\documentclass[aps,prb,twocolumn,reprint,superscriptaddress]{revtex4-2}
\usepackage{graphicx}
\usepackage{amsmath}
\usepackage{amssymb}
\usepackage{colordvi}
\usepackage{mathrsfs}
\usepackage{bm}
\usepackage{verbatim}
\usepackage{dcolumn}
\usepackage{epsfig}
\usepackage{subfigure}
\usepackage[colorlinks,allcolors=blue]{hyperref}
\usepackage{ulem}
\usepackage{dsfont}
\usepackage{makecell}
\usepackage{tikz}
\usepackage{lipsum,epsfig,dsfont}
\usepackage{booktabs}    
\usepackage{amsmath}     
\usepackage{subcaption}
\begin{document}
	\title{Topological Superconductors in Doubly-Coupled Nanowires with Altermagnetism}
	\author{Hongfa Pan}
	\affiliation{International Centre for Quantum Design of Functional Materials and Department of Physics, University of Science and Technology of China, Hefei, Anhui 230026, China}
	\affiliation{Hefei National Laboratory, University of Science and Technology of China, Hefei 230088, China}
	\author{Haoyang Wang}
	\affiliation{International Centre for Quantum Design of Functional Materials and Department of Physics, University of Science and Technology of China, Hefei, Anhui 230026, China}
	\author{Wenguang Zhu}
	\email[Correspondence author:~]{wgzhu@ustc.edu.cn}
	\affiliation{International Centre for Quantum Design of Functional Materials and Department of Physics, University of Science and Technology of China, Hefei, Anhui 230026, China}
	\affiliation{Hefei National Laboratory, University of Science and Technology of China, Hefei 230088, China}
	\author{Zhenhua Qiao}
	\email[Correspondence author:~]{qiao@ustc.edu.cn}
	\affiliation{International Centre for Quantum Design of Functional Materials and Department of Physics, University of Science and Technology of China, Hefei, Anhui 230026, China}
	\affiliation{Hefei National Laboratory, University of Science and Technology of China, Hefei 230088, China}
	\affiliation{Wilczek Quantum Center, Shanghai Institute for Advanced Studies, University of Science and Technology of China, Shanghai 201315, China}
	\date{\today{}}

\begin{abstract}
We theoretically investigate the possibility of engineering topological superconductivity in doubly-coupled nanowires integrating proximity-induced superconductivity and altermagnetism. By tuning experimentally accessible parameters, we find four distinct topological superconducting phases: class D, class BDI, and two phases hosting two Majorana zero modes per end, protected respectively by spin-group and magnetic point-group symmetries. Beyond inheriting the advantages of alternating-magnetism induced topological superconductivity, our system provides multiple tuning knobs (e.g., superconducting phase difference and inter-wire coupling) to control topological properties.
\end{abstract}

\maketitle
\section{Introduction} 
Topological superconductors (TSCs) hosting Majorana zero modes (MZMs) at their boundaries have attracted intense attention owing to their potential for fault-tolerant quantum computing~\cite{Kitaev2001,Green2000,C.Nayak2008,Alicea2011,M.Sato2016,Sato2017,HJGao2026}.
Despite decades of progress, there still exist three core challenges: developing experimentally feasible schemes for TSC realization, achieving unambiguous experimental identification of MZMs, and establishing protocols for controlled MZM manipulation~\cite{Alicea2011,rainis2012,Mourik2012,Furdyna2012,M.Sato2016,Sato2017,HJGao2026,kayyalha2020}. Single-nanowire platforms under external Zeeman fields, which have been extensively studied for topological superconductivity, show great promise~\cite{R.M.Lutchyn2010,Y.Oreg2010,C. M. Marcus2013,H. Shtrikma2012}.
However, strong Zeeman fields suppress the superconducting gap, and distinguishing MZMs from trivial Andreev bound states remains a persistent experimental difficulty~\cite{pan2020,yu2021,S. Das Sarma2023}. Altermagnetism, an emerging class of magnetic materials, has recently been a research focus in condensed matter physics~\cite{J. Liu2021,smejkal2022,J. LiuarXiv1,J. LiuarXiv2,smejkal2022b}. Unlike conventional Zeeman-field-based schemes that suppress the superconducting gap, altermagnetism enhances the gap significantly with zero net magnetization. Moreover, it allows for the use of weak external Zeeman field to distinguish MZMs from impurity-induced trivial states, representing a substantial advantage~\cite{E. Rossi2024}. Nevertheless, even with altermagnetism, single-nanowire platforms remain limited by the lack of MZM manipulation knobs~\cite{Sato2017,S. Das Sarma2023}.

Furthermore, the full experimental realization of TSCs with multiple symmetry-correlated boundary MZMs remains elusive due to pervasive technical constraints. Time-reversal-invariant TSCs hosting Majorana Kramers pairs, for instance, generally demand either unconventional superconducting substrates or prohibitively large values of an individual control parameter~\cite{keselman2013,zhang2013,gaidamauskas2014,C.L.M.Wong2012}, while topological crystalline superconductors require higher-dimensional platforms with dedicated crystalline symmetries~\cite{fang2014,J-FJ2024,S. Kobayashi}.

In this article, we theoretically investigate a doubly-coupled nanowire platform with proximity-induced superconductivity and altermagnetism from adjacent superconductors and altermagnets, respectively~\cite{E. Rossi2024}. Band degeneracy lifting by altermagnetism and Rashba spin-orbit coupling leads to the emergence of both intraband and interband pairing terms~\cite{M. Zehetmayer2013,E. Babaev2021,D. M. Paul2020,Belogolovskii2004,M. Amundsen2023,E. Blackburn2025}, whose interplay gives rise to class D TSC with one MZM per end and class BDI TSC with one or two MZMs per end, respectively~\cite{kitaev2009,schnyder2008,tewari2012,ryu2010,F.Pientka2017}. Notably, even within the BDI class, one MZM per end can be generically realized when the phase difference is not an integer multiple of $\pi$, making these phases directly applicable for topological quantum computing.
By introducing asymmetric altermagnetism or Rashba spin-orbit coupling between the two nanowires, we further find two topological phases protected respectively by spin-group symmetry~\cite{Q.-H. Liu2022,Jungwirt2022} and magnetic point-group symmetry, hosting two symmetry-related MZMs per end.
All identified topological phases occupy wide parameter ranges, with several continuously tunable through experimentally accessible controls, such as the superconducting phase difference and inter-wire coupling. This offers versatile tuning knobs for MZM manipulation and the detection of their hallmark experimental signatures.

The remainder of the paper is organized as follows. In Sec.~\ref{II}, the tight-binding model Hamiltonians are introduced. In Sec.~\ref{III}, the D class and BDI class TSCs are demonstrated. In Sec.~\ref{IV}, two types of TSCs are realized with two MZMs per end, protected by a spin-group symmetry and a magnetic point-group symmetry, respectively. In Sec.~\ref{V}, a brief conclusion is provided.

\section{Model and Hamiltonian}{\label{II}}
\begin{figure}
	\centering 
	\includegraphics[width=0.49\textwidth]{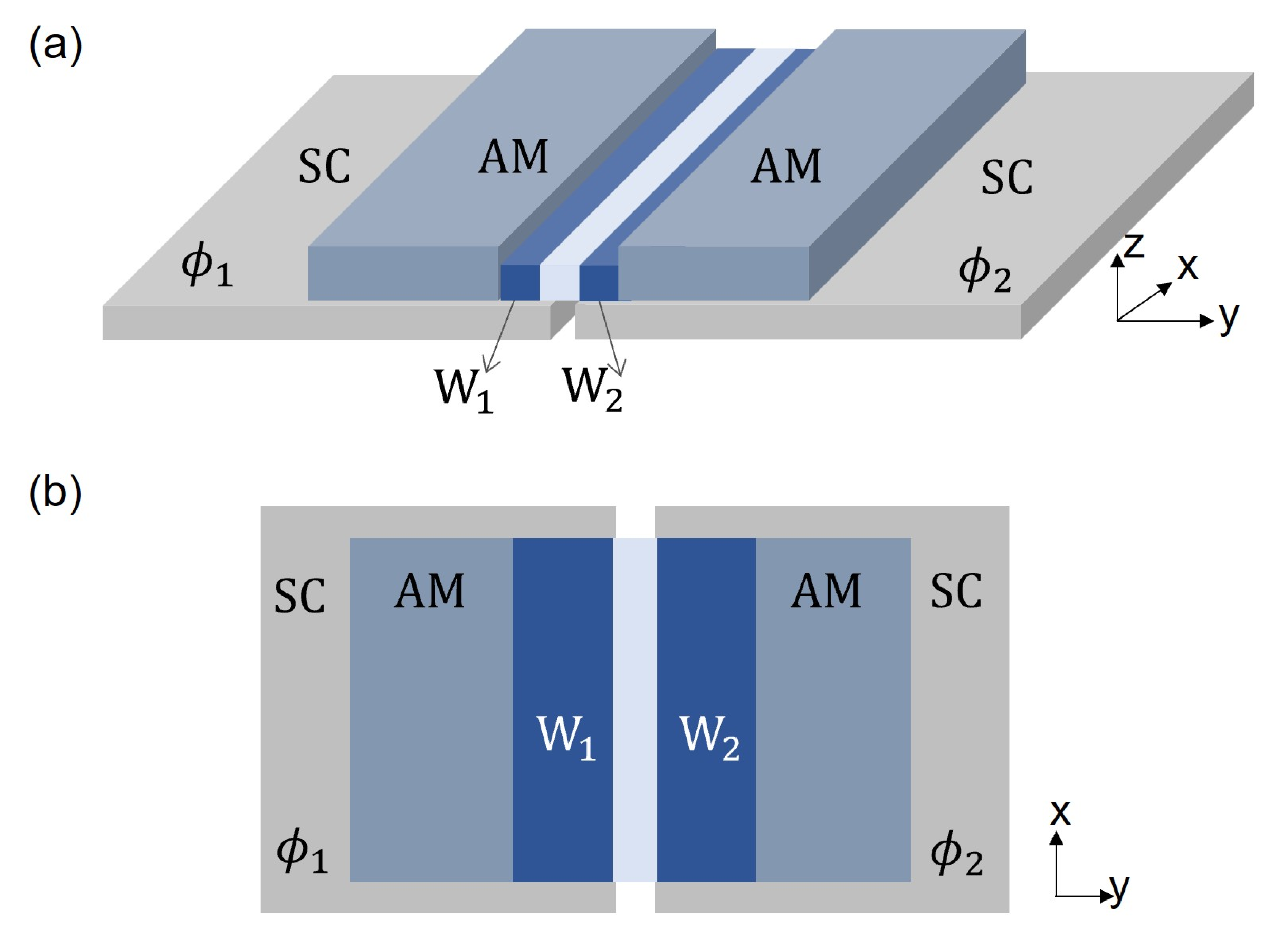}\caption{(a) Side-view and (b) top-view schematics of the doubly-coupled semiconducting nanowire platform with inter-wire coupling. Two superconducting (SC) regions exhibit superconducting phases $\phi_1$ and $\phi_2$, and two altermagnetic (AM) regions are labeled accordingly.}
	\label{fig1}.
\end{figure}

Figure~\ref{fig1} displays our system consisting of two coupled semiconductor nanowires with inter-wire coupling $t_0$, denoted as nanowires $w_{n}$ ($n=1,2$). $\mu$, $t$, and $\alpha_n$ denote respectively the chemical potential, nearest-neighbor hopping amplitude and Rashba spin-orbit coupling in the nanowire $w_{n}$. By proximity-coupled to altermagnets and superconductors, altermagnetic exchange coupling $J_n$ with $d_{x^2-y^2}$ symmetry~\cite{E. Rossi2024} and $s$-wave superconducting pairing potential $\Delta_n$ with phase $\phi_n$ are induced in each nanowire. The model Hamiltonian of the system reads:
\begin{equation}
	\begin{aligned}
		H&=H_{w_{1}}+H_{w_{2}}+H_{int},\\
		H_{w_{n}}&=H_{t_{n}}+H_{SO_{n}}+H_{J_{n}}+H_{\Delta_{n}},\\
		H_{t_{n}}&=-\mu\sum_{j,\alpha}^{}c^{\dagger}_{w_{n},j,\alpha}c_{w_{n},j,\alpha}\\
		&-t\sum_{j, \alpha}(c_{w_{n},j+1,\alpha}^{\dagger} c_{w_{n},j,\alpha}+\text {H.c.}),\\
		H_{SO_{n}}&=i\alpha_{n}\sum_{j,\alpha,\beta}(c_{w_{n},j+1, \alpha}^{\dagger}(\sigma_{y})_{\alpha,\beta}c_{w_{n},j,\beta}-\text{ H.c.}),\\
		H_{J_{n}}&=J_{n}\sum_{j,\alpha}^{}(c_{w_{n},j+1,\alpha}^{\dagger}\sigma_{z}c_{w_{n},j,\alpha}+\text{ H.c.}),\\
		H_{\Delta_{n}}&=\sum_j(\Delta_{n}e^{i\phi_{n}}c_{w_{n},j,\uparrow}^{\dagger}c_{w_{n},j, \downarrow}^{\dagger}+\text{H.c.}),\\
		H_{int}&=-t_{0}\sum_{j,\alpha}^{}(c^{\dagger}_{w_{1},j,\alpha}c_{w_{2},j,\alpha}+\text{H.c.}),
	\end{aligned}
\end{equation}
where $c^\dagger_{w_{n},j,\alpha}$ ($c_{w_{n},j,\alpha}$) creates (annihilates) an electron with spin $\alpha=\uparrow,\downarrow$ at site $j$ in nanowire $w_{n}$. Since only the relative phase difference between two superconductors is physically meaningful, we define the phases as $\phi_n = (-1)^{n}\phi/2$, where $\phi = \phi_2 - \phi_1$ is the phase difference across the two nanowires.
\begin{figure}
\centering 
\includegraphics[width=0.49\textwidth]{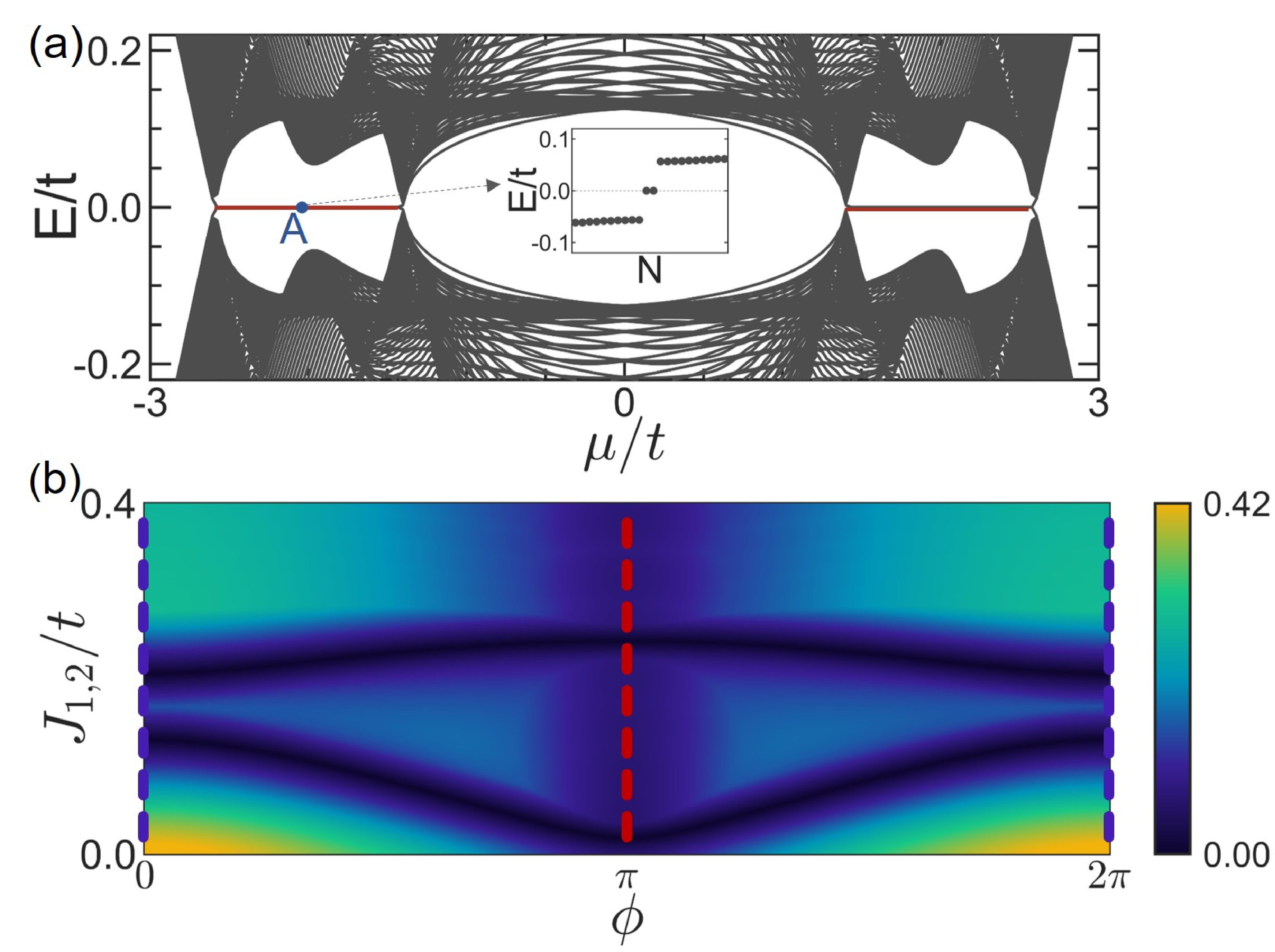}\caption{(a) Energy eigenvalues as a function of chemical potential $\mu$. Point A marks a twofold-degenerate zero-energy state, with an inset showing the eigenvalue distribution near zero energy. (b) Phase diagram in the plane of phase difference $\phi$ vs altermagnetism $J_{1,2}$.  Parameters are set to be $t=1$, $\alpha_{1,2}=0.35$, $\Delta_1=0.3$, $\Delta_2=0.15$, $t_0=0.25$. Additional parameters: (a) $J_{1,2}=0.2$, $\phi=0.6\pi$, and nanowire length $N_w=200$; (b) $\mu=-2$. }   
\label{fig2}.
\end{figure}

Since the momentum $k_x$ is conserved, the Hamiltonian $H$ can be transformed to momentum space. On the Nambu basis of $\{\psi^{\dagger}_{1},\psi_{1},\psi^{\dagger}_{2},\psi_{2}\}$, where the basis is defined as $\{\psi^{\dagger}_{n},\psi_{n}\}=\{c^{\dagger}_{n,\uparrow},c^{\dagger}_{n,\downarrow}, c_{n,\uparrow},c_{n,\downarrow}\}$, the corresponding Bogoliubov–de Gennes Hamiltonian takes the form
\begin{equation}
	\begin{aligned}
	H_{k}&=H_{k1}\frac{s_{0}+s_{z}}{2}+H_{k2}\frac{s_{0}-s_{z}}{2}-t_{0}s_{x}\tau_{z},\\
	H_{kn}&=-(\mu+2t\cos{k_{x}})\tau_{z}+2J_{n}\cos{k_{x}}\tau_{z}\sigma_{z}\\&+2\alpha_{n}\sin{k_{x}}\tau_{z}\sigma_{y}+e^{i\phi_{n}}\Delta_{n}\frac{\tau_{x}+i\tau_{y}}{2}i\sigma_{y}\\&-e^{-i\phi_{n}}\Delta_{n}\frac{\tau_{x}-i\tau_{y}}{2}i\sigma_{y},\\
	\end{aligned}\label{equation.2}
\end{equation}
where $s,\tau,\sigma$ are Pauli matrices acting on the the wire, particle-hole, and spin spaces, respectively. The particle-hole symmetry of the Hamiltonian $H_{k}$ is characterized by the operator $P=\tau_{x}K$, where $K$ denotes complex conjugation. 

The double-wire architecture endows the system with a significantly enlarged Hilbert space compared to single-wire counterpart, introducing additional wire-specific degrees of freedom that expand the parameter space accessible to experimental manipulation. This rich parameter space enables control over the key symmetry operations: the mirror symmetry exchanging the two wires, as well as its nontrivial combinations with particle-hole symmetry and time-reversal symmetry, can be either preserved or broken in different parameter regimes. For instance, the wire-exchange mirror symmetry is broken when the superconducting pairing amplitudes of the two wires are unequal ($\Delta_1 \neq \Delta_2$), placing the system in the D class as a single TSC nanowire when their relative phase $\phi$ is not an integer multiple of $\pi$ ($\phi \neq 0,\pi,2\pi$). In contrast, when $\phi = 0$, the system realizes the BDI symmetry class, which constitutes a paradigmatic example of coupled TSC nanowire systems~\cite{tewari2012}. Remarkably, even when $\phi$ deviates from integer multiples of $\pi$, the underlying symmetries can interconnect the two wires and restore the BDI symmetry class. These key symmetries enable the design of distinct TSCs as shown in Table~\ref{tab:1} and Table~\ref{tab:2}.

\begin{table}[htbp]
	\centering
	\caption{Topological classes in the Altland–Zirnbauer tenfold classification. All unlisted parameters are identical for the two nanowires in all cases.}
	\label{tab:1}
	\renewcommand{\arraystretch}{1.3}
	\begin{tabular}{ccc}
		\toprule
		$\Delta_1 = \Delta_2$ & $\phi$ & Class \\
		\midrule
		Yes &  arbitrary & BDI \\
		No & $\phi=m\pi$ & BDI \\
		No & $\phi \neq m\pi$ & D \\
		\bottomrule
	\end{tabular}
\end{table}
\begin{table}[htbp]
	\centering
	\caption{The case of MZMs related by detailed symmetry. All unlisted identical parameters are equal for the two nanowires in all cases.}
	\label{tab:2}
	\renewcommand{\arraystretch}{1.3}
	\begin{tabular}{cccc}
		\toprule
	 $J_1=\pm J_2$ & $\phi$ & $\alpha_1=\pm\alpha_2$ & Group \\
	 	\midrule
$-$ & $0$ & $+$ &  $T_{1}$ \\ 
$-$ & $\pi$ & $-$ & $T_{2}$ \\
		\bottomrule
	\end{tabular}
\end{table}
\begin{figure*}
	\centering 
	\includegraphics[width=0.98\textwidth]{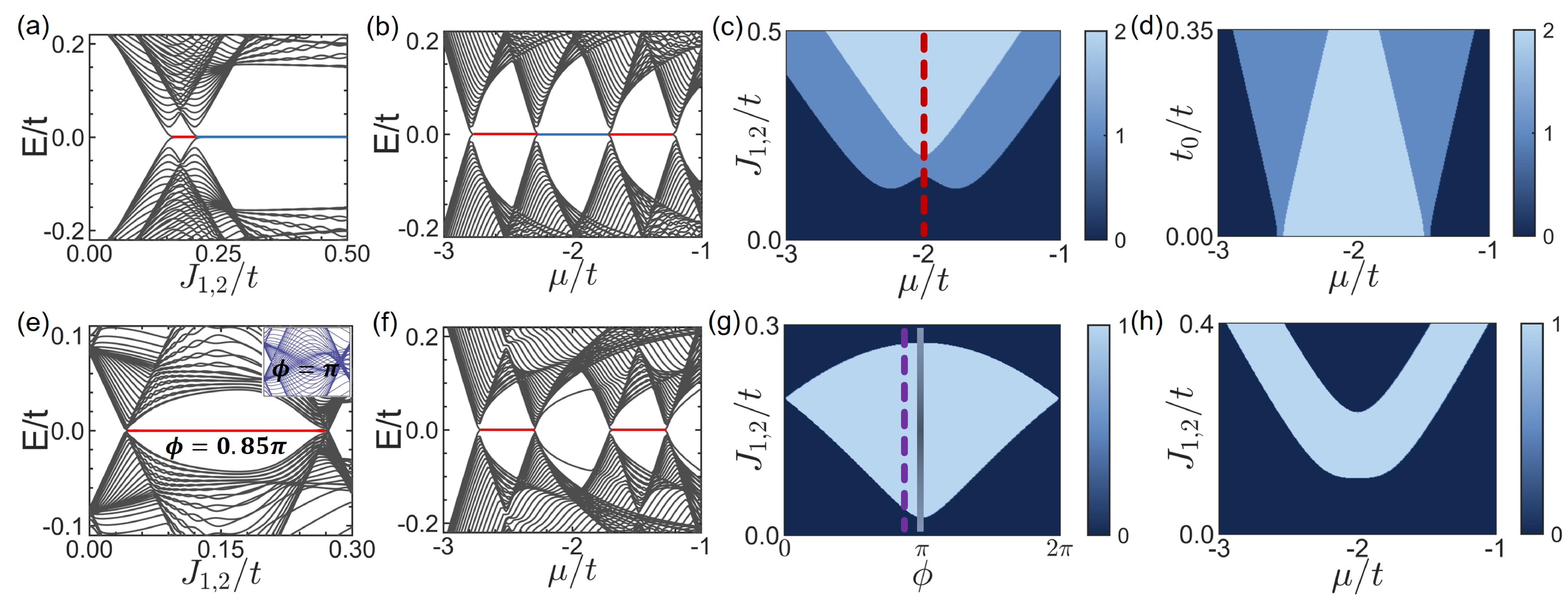}\caption{(a)-(d) BDI TSCs with $\phi=0$: (a) Energy eigenvalues as a function of altermagnetism $J_{1,2}$. (b) Energy eigenvalues as a function of the chemical potential $\mu$. (c) The phase diagram as the function of $\mu$ and $J_{1,2}$. (d) The phase diagram as the function of $\mu$ and $t_{0}$. Parameters used in all calculations: $t=1$, $\alpha_{1,2}=0.35$, $\Delta_1=0.3$, $\Delta_2=0.2$. Additional parameters: (a) $\mu=-2$, $t_0=0.25$ and nanowire length $N_{w}=100$. (b) $J_{1,2}=0.3$, $t_0=0.25$ and nanowire length $N_{w}=100$. (c) $t_{0}=0.25$; (d) $J_{1,2}=0.3$.
		(e)-(h) BDI TSCs with $\phi\neq m\pi$ ($m=0,1,2$): (e) Energy eigenvalues as a function of the altermagnetism $J_{1,2}$. (f) Energy eigenvalues as a function of the chemical potential $\mu$. (g) Phase diagram as a function of $\phi$ and $J_{1,2}$. (h) Phase diagram as the function of $\mu$ and $J_{1,2}$. Parameters used in all calculations: $t=1$, $\alpha_{1,2}=0.3$, $\Delta_{1,2}=0.3$. Additional parameters: (e) $\mu=-2$, $t_{0}=0.25$, $\phi=0.85\pi,\pi$ and nanowire length $N_{w}=100$. (f) $J_{1,2}=0.3$, $t_{0}=0.2$, $\phi=0.5\pi$ and nanowire length $N_{w}=100$. (g) $\mu=-2$ and $t_{0}=0.25$; 
		(h) $t_{0}=0.2$ and $\phi=0.5\pi$.}   
	\label{fig3}.
\end{figure*}

\section{D class and BDI class Topological superconductors} \label{III}
By setting $J_1=J_2$ and $\alpha_1=\alpha_2$, the system can be reduced to a two-band superconductor with both intraband and interband pairings [See Appendix.~\ref{A2}], whose amplitudes are respectively $\Delta_s$ and $\Delta_d$,
\begin{equation}
\begin{aligned}	
\Delta_{s}&=|\frac{1}{2} e^{-\frac{i\phi}{2}}(\Delta_{1}+\Delta _{2}e^{i\phi})|, \\
\Delta_{d}&=|\frac{1}{2} e^{-\frac{i\phi}{2}}(\Delta_{2}e^{i\phi}-\Delta_{1})|.
\end{aligned}
\end{equation}

In this parameter regime, the topological phases are governed by the interplay between interband pairing $\Delta_d$ and intraband pairing $\Delta_s$, where the latter corresponds to the pairing mechanism in conventional single-nanowire TSCs. This results in a rich topological phase diagram with multiple independently tunable parameters.

\subsection{D class topological superconductors}
When both the time-reversal symmetry is broken by the altermagnetism and additional symmetries are lifted by imposing $|\Delta_1| \neq |\Delta_2|$ and $\phi \neq m\pi$ for integer $m$, the system is driven into the D class. The conditions for topologically nontrivial states are signaled by the bulk gap closing at momenta $k_x=0$ or $k_x=\pi$~\cite{Kitaev2001,E. Rossi2024,F.Pientka2017}. The topological phase boundary depends on a variety of experimentally tunable parameters [See Appendix.~\ref{A5}], including the phase difference $\phi$, inter-wire coupling $t_0$, chemical potential $\mu$, and altermagnetism $J_{1,2}$. As a consequence, the topological properties can be modulated via multiple independent knobs. Specifically, the evolution of energy spectrum with chemical potential $\mu$ is displayed in Fig.~\ref{fig2}(a). Zero-energy modes exist within a relatively broad range of $\mu$. For example, point A corresponds to the presence of two MZMs, which is the generic case for D class TSCs. The altermagnetism $J_{1,2}$ and phase difference $\phi$ also provide powerful knobs. The phase diagram as a function of $\phi$ and $J_{1,2}$ is shown in Fig.~\ref{fig2}(b). The region enclosed by the two dark blue curves, which mark the gap-closing topological phase transitions, is topologically nontrivial. This demonstrates an effective approach in engineering topological superconductivity via the phase difference $\phi$, which relaxes the requirement of strong altermagnetism for topological phase when $\phi$ becomes finite.

\subsection{BDI class topological superconductors}
At some special points of $\phi=m\pi$ ($m=0,1,2$), the system belongs to the BDI class, denoted by the blue dashed lines and red dashed line in Fig.~\ref{fig2}(b). When $\phi=0,2\pi$ ($\phi=\pi$), the Hamiltonian $H_k$ possesses an anti-unitary time-reversal symmetry $\mathcal{T}_1=K$ ($\mathcal{T}_2=\tau_z K$), which together with the particle-hole symmetry $P$ generates a unitary chiral symmetry $S_1=P\mathcal{T}_1$ ($S_2=P\mathcal{T}_2$) satisfying $\{S_i, H_k\}=0$ for $i=1,2$. Thus, the system falls into the BDI class with $\mathbb{Z}$ topological invariant, which counts the number of MZMs localized per end of the nanowire. As displayed in Appendix~\ref{A2}, the interplay between two pairing channels gives rise to two distinct types of BDI class topological superconductors.

At $\phi=\pi$, the small bulk gap is dominated by the intraband pairing $\Delta_s=|\frac{1}{2}(\Delta_1-\Delta_2)|$, which is proportional to the difference between two pairings, as illustrated in Fig.~\ref{fig2}(a). In contrast, at $\phi=0$, the gap is governed by $\Delta_s=|\frac{1}{2}(\Delta_1+\Delta_2)|$, leading to a significantly larger bulk gap. Hereinbelow, we focus on the $\phi=0$ case. With the increase of altermagnetism $J_{1,2}$, two consecutive topological phase transitions occur, marked by the emergence of zero-energy modes corresponding to topologically nontrivial phases with one and two MZMs per end of the doubly coupled nanowires, respectively [see Fig.~\ref{fig3}(a)]. Similarly, the variation of chemical potential $\mu$ drives two transitions to zero-energy phases, followed by two transitions back to trivial phases, corresponding sequentially to phases with one MZM, two MZMs, and no MZMs per end [Fig.~\ref{fig3}(b)]. To systematically explore the parameter dependence of the topological phases, we present phase diagrams as functions of chemical potential $\mu$ versus altermagnetism $J_{1,2}$ and inter-wire coupling $t_0$ in Figs.~\ref{fig3}(c) and \ref{fig3}(d), respectively. Over a broad parameter range, we find topologically nontrivial phases with $\mathbb{Z}=1$ and $\mathbb{Z}=2$, corresponding to one and two MZMs per end, respectively. The red line in Fig.~\ref{fig3}(c) represents the parameter cut shown in Fig.~\ref{fig3}(a), which traverses the trivial phase, $\mathbb{Z}=1$ nontrivial phase, and $\mathbb{Z}=2$ nontrivial phase.

Remarkably, even when $\phi \neq m\pi$, the system remains in the BDI class for $\Delta_1 = \Delta_2$. In this parameter regime, an effective anti-unitary time-reversal symmetry $\mathcal{T}_3 = M K$ emerges, where $M = s_x$ denotes the operator exchanging the two nanowires. This effective time-reversal symmetry combines with the particle-hole symmetry $P$ to generate a unitary chiral symmetry $S_3 = P \mathcal{T}_3$, satisfying $\{S_3, H_k\} = 0$. The evolution of the energy spectrum with altermagnetism $J_{1,2}$ and chemical potential $\mu$ is presented in Figs.~\ref{fig3}(e) and \ref{fig3}(f), respectively, revealing topologically nontrivial phases over a broad parameter range. As displayed in Fig.~\ref{fig3}(e), at $\phi = 0.85\pi$, zero-energy modes corresponding to MZMs persist over a wide range of $J_{1,2}$. Near $\phi = \pi$, the bulk gap tends to close because the intraband pairing amplitude $\Delta_s = |\frac{1}{2}(\Delta_1 - \Delta_2)|$ vanishes when $\Delta_1 = \Delta_2$. These two cases are respectively marked by the purple and blue lines in Fig.~\ref{fig3}(g), which shows the phase diagram as a function of phase difference $\phi$ and altermagnetism $J_{1,2}$.
Away from $\phi = \pi$ point, the system exhibits a topologically nontrivial phase with one MZM per end over a wide range of $J_{1,2}$ and $\phi$, with a strong dependence on the phase difference $\phi$, which also relaxes the requirement for strong altermagnetism. Topologically nontrivial phases over a broad parameter range is also observed in the phase diagram as a function of $\mu$ and $J_{1,2}$ in Fig.~\ref{fig3}(h).
\begin{figure*}
	\centering 
	\includegraphics[width=0.98\textwidth]{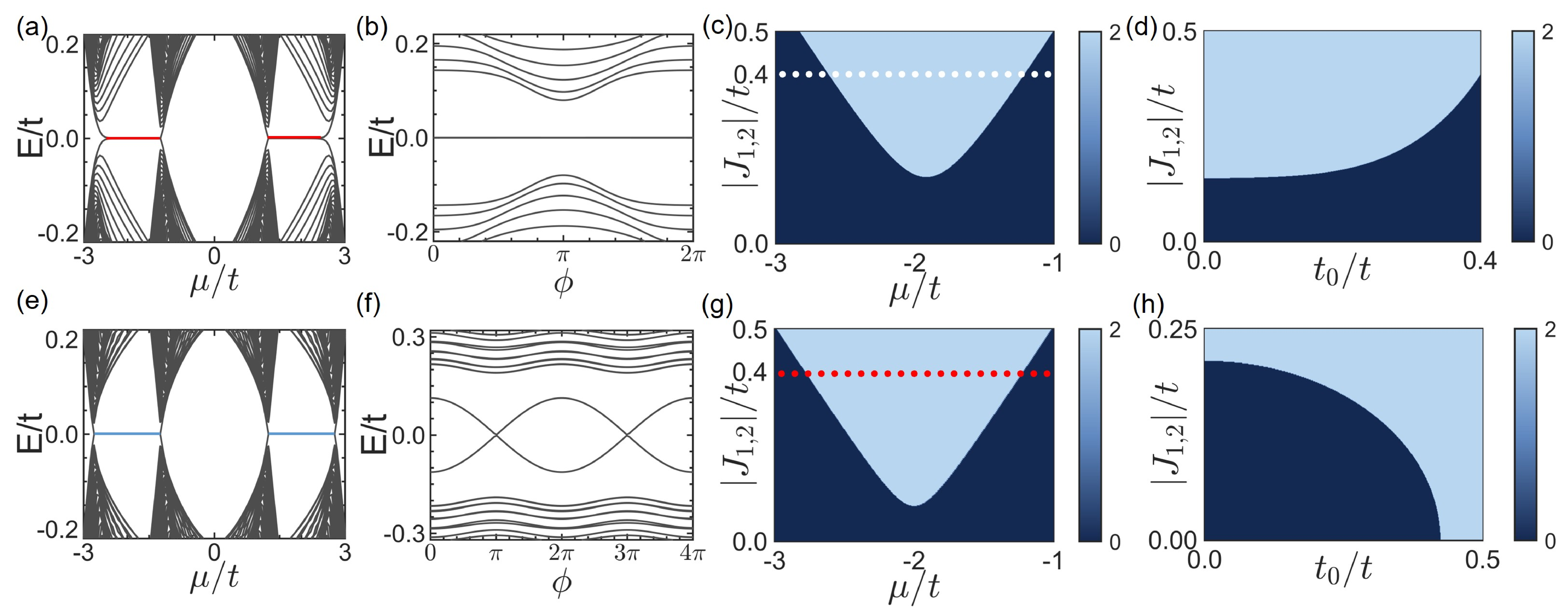}\caption{(a)-(d) spin-group symmetry protected TSCs: (a) Energy eigenvalues as a function of the chemical potential $\mu$. (b) Energy eigenvalues as a function of the phase difference $\phi$. (c) Phase diagram as a function of $\mu$ and $J_{1,2}$. (d) Phase diagram as a function of $t_{0}$ and $J_{1,2}$. Parameters used in all calculations: $t=1$ and $\Delta_{1,2}=0.3$. Additional parameters: (a) $\alpha_{1,2}=0.35$, $J_{1,2}=\pm0.4$, $t_{0}=0.2$, $\phi=0$ and nanowire length $N_{w}=100$; (b) $\mu=-2$ and the other fixed parameters are same as those in (a); (c) $J_{1}=-J_{2}$ and the fixed parameters are same as those in (a); (d) $J_{1}=-J_{2}$, $\alpha_{1,2}=0.3$, $\mu=-2$ and $\phi=0$. (e)-(h) Magnetic point-group symmetry protected TSCs:  (e) Energy eigenvalues as a function of the chemical potential $\mu$. (f) Energy eigenvalues as a function of the phase difference $\phi$. (g) Phase diagram as a function of $\mu$ and $J_{1,2}$. (h) Phase diagram as a function of $t_{0}$ and $J_{1,2}$. Parameters used in all calculations: $t=1$, $\alpha_{1,2}=\pm0.3$, and $\Delta_{1,2}=0.3$. Additional parameters: (e) $J_{1,2}=\pm0.4$, $t_{0}=0.25$, $\phi=\pi$ and nanowire length $N_{w}=100$; (f) $\mu=-2$ and the other fixed parameters are same as those in (e); (g) $J_{1}=-J_{2}$ and the fixed parameters are same as those in (e); (h) $J_{1}=-J_{2}$, $\mu=-2.3$ and the other fixed parameters are same as those in (e).}   
	\label{fig4}.
\end{figure*}
\section{spin-group symmetry and magnetic point-group symmetry protected Topological superconductors} \label{IV}
TSCs protected by magnetic point-group symmetry or time-reversal symmetry have been extensively studied both theoretically and experimentally, where multiple MZMs can emerge at system boundaries. However, most of these realizations are based on higher-dimensional systems or unconventional pairings~\cite{fang2014,J-FJ2024,S. Kobayashi}. In this Section, we demonstrate that two distinct types of TSCs can be realized in our quasi-one-dimensional system by tuning either the exchange couplings $J_{1,2}$ or the Rashba spin-orbit couplings $\alpha_{1,2}$. These two phases are protected by spin-group symmetry~\cite{Q.-H. Liu2022,Jungwirt2022} and magnetic point-group symmetry, respectively.

In the first case, By setting $J_1=-J_2$, $\phi=0$, and all other parameters to be identical in the two nanowires, the anti-unitary spin-group symmetry $T_1=M\mathcal{T}$ is preserved, satisfying $T_1 H_k(k_x) T_1^{-1}=H_k(-k_x)$. $\mathcal{T}=i\sigma_y K$ is the time-reversal operator. In the topologically nontrivial phase, two MZMs reside per end of the coupled nanowire system, related by the $T_1$ symmetry. The nontrivial phase is denoted by the red lines in Fig.~\ref{fig4}(a). Since this symmetry is composed of $M$ and $\mathcal{T}$, and cannot be broken by a phase difference, a nonzero $\phi$ does not destroy the topologically nontrivial phase, in contrast to the time-reversal-invariant TSCs protected solely by $\mathcal{T}$, as illustrated in Fig.~\ref{fig4}(b). To characterize this topological invariant, we introduce a new chiral symmetry $S_{T_1}=PT_1$ under these parameter settings. The combination with the chiral symmetry $S_1=P\mathcal{T}_{1}$ yields a unitary symmetry $U_1=S_1 S_{T_1}$ that commutes with the Hamiltonian, $[U_1,H_k]=0$. By block-diagonalizing the Hamiltonian using $U_1$ symmetry, the integer topological invariant $\mathbb{Z}$ of the BDI class can be computed for each block via the $S_1$ chiral symmetry. The phase diagrams as functions of $\mu$ vs $J_{1,2}$ and $t_0$ vs $J_{1,2}$ are presented in Figs.~\ref{fig4}(c) and \ref{fig4}(d), respectively. A broad topologically nontrivial region hosting two MZMs per end, corresponding to $\mathbb{Z}=2$, is clearly present in Fig.~\ref{fig4}(c). The white line in Fig.~\ref{fig4}(c) corresponds to the parameter trajectory shown in Fig.~\ref{fig4}(a). Figure~\ref{fig4}(d) further shows that the critical altermagnetism $J_{1,2}$ required for the topologically nontrivial phase increases with increasing $t_0$.

The second case is characterized by the parameter relations $J_1=-J_2$, $\alpha_1=-\alpha_2$, and $\phi=\pi$, with all other parameters being identical in the two nanowires. A distinct anti-unitary symmetry $T_2=\widetilde{M}\mathcal{T}$ of magnetic point-group symmetry is preserved, satisfying $T_2 H_k(k_x) T_2^{-1}=H_k(-k_x)$, where $\widetilde{M}=s_x\tau_z\sigma_z$ and $\mathcal{T}=i\sigma_y K$.
As shown in Fig.~\ref{fig4}(e), the nontrivial phase is denoted by the blue lines. The energy spectrum exhibits a distinct $4\pi$ periodicity as a function of phase difference $\phi$ (see Fig.~\ref{fig4}(f)), which provides a clear experimental signature for this TSC phase. A unitary symmetry $U_2=S_{T_2} S_2$ that commutes with the Hamiltonian, $[U_2,H_k]=0$, can also be constructed, where $S_{T_2}=PT_2$ and $S_2=P\mathcal{T}_2$ are chiral symmetries with $\mathcal{T}_2=\tau_z K$. The integer topological invariant $\mathbb{Z}$ can be computed similarly via the $S_2$ chiral symmetry.
The phase diagrams as functions of $\mu$ vs $J_{1,2}$ and $t_0$ vs $J_{1,2}$ are displayed in Figs.~\ref{fig4}(g) and \ref{fig4}(h), respectively.
A broad region hosting two MZMs, corresponding to $\mathbb{Z}=2$, is shown in Figs.~\ref{fig4}(g) and \ref{fig4}(h). The red line in Fig.~\ref{fig4}(g) corresponds to the parameter trajectory shown in Fig.~\ref{fig4}(e). It also relaxes the requirement for altermagnetism or the inter-wire coupling when the other is non-zero as shown in Fig.~\ref{fig4}(h).
\section{conclusion}\label{V}
To conclude, our work extends the frontier of altermagnetism-based topological superconductivity beyond single nanowires, introducing a versatile platform with unparalleled experimental tunability. This system hosts the D class and BDI class topological superconducting phases, and phases protected by spin-group and magnetic point-group symmetries. The rich set of independent tuning knobs inherent to this architecture may enable the detection of topological superconducting signatures and implementation of non-Abelian braiding and fusion of MZMs.

\section*{ACKNOWLEDGMENTS} 	
This work was financially supported by Innovation Program for Quantum Science and Technology (Grant No. 382
2021ZD0302800) and the National Natural Science Foundation of China (Grants No. 380
12474158, No. 12234017, and No. 12488101). We also thank the Supercomputing Center of University of Science and Technology of China for providing high-performance computing resources.

\appendix
\section{}
\subsection{Model in the D class and BDI class TSCs}
For both D class and BDI class TSCs, the Hamiltonian $H_k$ can be transformed to a new Hamiltonian $H_d$ expressed in the basis $\{\psi^\dagger_{d1}, \psi_{d1}, \psi^\dagger_{d2}, \psi_{d2}\}$, where
\begin{equation}
	\begin{aligned}
	\psi^{\dagger}_{d1}&=\frac{1}{\sqrt{2}}\{c^{\dagger}_{2,\uparrow}-c^{\dagger}_{1,\uparrow},c^{\dagger}_{2,\downarrow}-c^{\dagger}_{1,\downarrow}\},\\
	\psi^{\dagger}_{d2}&=\frac{1}{\sqrt{2}}\{c^{\dagger}_{2,\uparrow}+c^{\dagger}_{1,\uparrow},c^{\dagger}_{2,\downarrow}+c^{\dagger}_{1,\downarrow}\}
	\end{aligned}
\end{equation}
The Hamiltonian $H_{d}$ can be denoted by 
\begin{equation}
H_{d}(k)=\begin{pmatrix}
H_{1} & H_{\Delta_{12}}\\ 
H^{\dagger}_{\Delta_{12}} &H_{2} \label{A2}
    \end{pmatrix}
\end{equation}
where the sub-matrices $H_{i}$ with $i=1,2$ and $H_{\Delta_{12}}$ are denoted respectively by 
\begin{small}
	\begin{widetext}
		\begin{equation}
			H_{i}=\begin{pmatrix}
				2(J_{1}-t)\cos(k)-\mu_{i} &-2i\alpha_{1}&0&\frac{1}{2} e^{-\frac{i\phi}{2}}(\Delta_{1}+\Delta _{2}e^{i\phi})\\ 
				2i\alpha_{1}&2(-J_{1}-t)\cos(k)-\mu_{i}&-\frac{1}{2} e^{-\frac{i\phi}{2}}(\Delta_{1}+\Delta _{2}e^{i\phi})&0\\
				0&-\frac{1}{2}e^{-\frac{i\phi}{2}}(\Delta_{2}+\Delta _{1}e^{i\phi})&2(-J_{1}+t)\cos(k)+\mu_{i}& 2i\alpha_{1}\\
				\frac{1}{2} e^{-\frac{i\phi}{2}}(\Delta_{2}+\Delta _{1}e^{i\phi})&0& -2i\alpha_{1}&2(J_{1}+t)\cos(k)+\mu_{i}\\
			\end{pmatrix}
		\end{equation}
	\end{widetext}
\end{small}

and
\begin{small}
	\begin{widetext}
		\begin{equation}
			H_{\Delta_{12}}=\begin{pmatrix}
				0 & 0&0&\frac{1}{2} e^{-\frac{i\phi}{2}}(-\Delta_{1}+\Delta_{2}e^{i\phi})\\ 
			    0&0&\frac{1}{2} e^{-\frac{i\phi}{2}}(\Delta_{1}-\Delta_{2}e^{i\phi})&0\\
				0&\frac{1}{2} e^{-\frac{i\phi}{2}}(\Delta_{1}e^{i\phi}-\Delta_{2})&0& 0\\
				\frac{1}{2}e^{-\frac{i\phi}{2}}(-\Delta_{1}e^{i\phi}+\Delta_{2})&0& 0&0\\
			\end{pmatrix}
		\end{equation}
	\end{widetext}
\end{small}
where $\mu_{1}=-t_{0}+\mu$ and $\mu_{2}=t_{0}+\mu$.

Thus, the Hamiltonian describes a two-band superconductor with both intraband and interband pairing terms~\cite{M. Zehetmayer2013,E. Babaev2021,D. M. Paul2020,Belogolovskii2004,M. Amundsen2023,E. Blackburn2025}, characterized by an effective chemical potential difference $2t_{0}$.
Intraband pairing favors the formation of a TSC state, whereas interband pairing tends to suppress it~\cite{tewari2012,Kitaev2001}. The magnitudes of these two pairing components are given by
\begin{equation}
	\Delta_{s}=|\frac{1}{2} e^{-\frac{i\phi}{2}}(\Delta_{1}+\Delta _{2}e^{i\phi})|
\end{equation}
\begin{equation}
	\Delta_{d}=|\frac{1}{2} e^{-\frac{i\phi}{2}}(\Delta_{2}e^{i\phi}-\Delta_{1})|
\end{equation}
which can be controlled by tuning the phase difference $\phi$ and the pairing strengths $\Delta_1$ and $\Delta_2$.

When $\Delta_{1}=\Delta_{2}$ and $\phi=0,2\pi$, The magnitudes of these two pairing components reduce to
\begin{equation}
	\Delta_{s}=|\Delta_{1}|, \quad \Delta_{d}=0
\end{equation}
Only intraband pairing dominates, allowing the model to be mapped onto two decoupled Kitaev p-wave superconducting chains with effective chemical potentials $t_{0}-\mu$ and $-t_{0}-\mu$, respectively. As a special case, this model belongs to the BDI symmetry class~\cite{tewari2012}.

When $\Delta_{1}=\Delta_{2}$ and $\phi=\pi$, the magnitudes of these two pairing components reduce to
\begin{equation}
	\Delta_{s}=0, \quad \Delta_{d}=|i\Delta_{1}|
\end{equation}
Only interband pairing dominates in this regime. When the $t_{0}$ and $-t_{0}$ chemical potential shifts are omitted, the model can similarly be mapped onto two decoupled Kitaev p-wave superconducting chains. In the general case, these shifts globally displace the energies of the bands originating from the pairing terms. When $t_{0}$ is comparable to the pairing amplitude, as shown in Fig.~\ref{fig3}(c), the superconducting gap typically closes.

When $\Delta_{1}\neq\Delta_{2}$ and $\phi=0,2\pi$, the magnitudes of these two pairing components reduce to
\begin{equation}
	\Delta_{s}=\frac{1}{2}|\Delta_{1}+\Delta_{2}|, \Delta_{d}=\frac{1}{2}|\Delta_{1}-\Delta_{2}|
\end{equation}
It is obvious that the intraband pairing dominates when the difference between $\Delta_{1}$ and $\Delta_{2}$ is small. 

For $\Delta_{1}\neq\Delta_{2}$ with $\phi=\pi$, the magnitudes of these two pairing components reduce to
\begin{equation}
	\Delta_{s}=|\frac{i}{2}(\Delta_{1}-\Delta_{2})|, \Delta_{d}=|\frac{i}{2}(\Delta_{1}+\Delta_{2})|
\end{equation}
The intraband pairing is dependent on the difference between $\Delta_{1}$ and $\Delta_{2}$.
 
\subsection{Conditions of D class TSCs}
The conditions for the topologically nontrivial phase are determined by the closing of the bulk energy gap in the Hamiltonian $H_k$ at $k_x=0$ or $\pi$~\cite{Kitaev2001,E. Rossi2024}.

At $k_x=0$, the critical exchange coupling for gap closure is given by
\begin{equation}
	J_{c1}=\frac{\sqrt{m_0\pm\sqrt{n_0}}}{2\sqrt{2}},
	\label{A5}
\end{equation}
with $m_0=2(2t+\mu)^2+\Delta_1^2+\Delta_2^2+2t_0^2$ and $n_0=16t_0^2(2t+\mu)^2+(\Delta_1^2-\Delta_2^2)^2+4t_0^2(\Delta_1^2+\Delta_2^2-2\Delta_1\Delta_2\cos\phi)$.

At $k_x=\pi$, the corresponding critical exchange coupling reads
\begin{equation}
	J_{c2}=\frac{\sqrt{m_\pi\pm\sqrt{n_\pi}}}{2\sqrt{2}},
	\label{A6}
\end{equation}
with $m_\pi=2(-2t+\mu)^2+\Delta_1^2+\Delta_2^2+2t_0^2$ and $n_\pi=16t_0^2(-2t+\mu)^2+(\Delta_1^2-\Delta_2^2)^2+4t_0^2(\Delta_1^2+\Delta_2^2-2\Delta_1\Delta_2\cos\phi)$.

The topological phase boundary is thus defined by Eqs.~\eqref{A5} and \eqref{A6}, which is dependent on $\mu$. For example with $\mu\times t<0$, the system resides in the topologically nontrivial phase with $\frac{\sqrt{m_0-\sqrt{n_0}}}{2\sqrt{2}}<J_{1,2}<\frac{\sqrt{m_0+\sqrt{n_0}}}{2\sqrt{2}}$.
These critical conditions depend on a variety of experimentally adjustable parameters.
In particular, the topologically nontrivial phase can be effectively controlled by the phase difference $\phi$ when $|t|$ is comparable to $|\mu|$, since all other parameters are generally small in realistic experimental settings. In this regime, $J_{c1}$ is dominated by $\Delta_{1,2}$ and $t_0$ of comparable magnitude, can be continuously tuned via $\phi$, as shown in Fig.~\ref{fig1}(d).
\subsection{Calculation of the topological invariant}
A chiral symmetry $S_0$ is defined as $\{S_0,H_0\}=0$.
In the basis where $S_0$ is diagonal, $H_0$ takes the block-off-diagonal form:
\begin{equation}
	H(k)=\begin{pmatrix}
		0 & A(k) \\
		A^T(-k) & 0
	\end{pmatrix}.
\end{equation}
The winding number $W$ is given by
\begin{equation}
	W = \frac{-i}{2\pi} \int_{-\pi}^{\pi} \frac{dz(k)}{z(k)},
\end{equation}
where the complex function $z(k)=\det[A(k)]/|\det[A(k)]|$.

For TSCs protected by spin-group symmetry and magnetic point-group symmetry in our system, the Hamiltonian $H_k$ can be block-diagonalized using the unitary symmetries $U_1$ and $U_2$, respectively, yielding
\begin{equation}
	H(k)=\begin{pmatrix}
		H_a & 0 \\
		0 & H_b
	\end{pmatrix}.
\end{equation}
The total topological invariant for these phases is obtained by summing the invariants of the two diagonal blocks, each of which is calculated using the chiral symmetry acting on the corresponding submatrix.

\end{document}